\documentclass[fleqn,twoside]{article}
\usepackage{espcrc2}

\usepackage{graphicx}
\usepackage[figuresright]{rotating}

\newcommand{\AmS}{{\protect\the\textfont2
  A\kern-.1667em\lower.5ex\hbox{M}\kern-.125emS}}

\title{R$\&$D efforts towards a neutrino factory}

\author{M. Bonesini\address{Sezione INFN Milano-Bicocca\\
Piazza Scienza 3, 20126-Milano, Italy}}
       
\begin{document}

\begin{abstract}
The R\&D efforts towards a neutrino factory are outlined with special emphasis 
on the muon cooling issue and the data collected for target optimization.
\end{abstract}

\maketitle

\section{Introduction.}
\begin{figure}
\includegraphics[width=\linewidth]{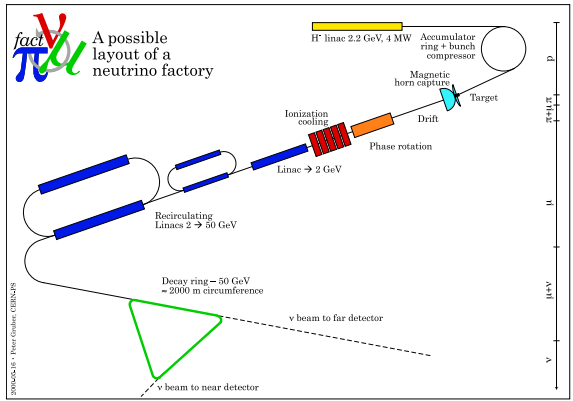}
\vskip -1cm
\caption{Schematic layout of a Neutrino Factory (CERN design).}
\label{nf}
\end{figure}

The neutrino factory ($\nu F$) is a muon storage ring   
where decaying muons produce collimated neutrino beams 
along its straight sections.
Several $\nu$F designs have been proposed, such as the ones of references
\cite{US2,cern}: 
the CERN design  is shown in Figure \ref{nf}.
A high intensity beam accelerated by a high power proton driver produces 
in a thin
Hg target, after some accumulation and bunch compression, low energy pions.
After a collection system, muons are cooled and phase rotated before acceleration
up to 20-50 GeV/c, depending on the design. Accelerated muons of well defined charge and momentum are then
injected into an accumulator where they circulate until they decay, giving two
neutrino beams along the straight sections.
The physics program at a neutrino factory is very rich and 
includes long-baseline $\nu$
oscillations, short-baseline $\nu$ physics and slow muon physics \cite{dydak}.
For the design of a $\nu$F some key points have to be clarified with
dedicated R\&D experiments. They include targetry, both MC validation and 
feasibility studies of the target-pion collection complex, $\mu$
cooling and accelerator R\&D (mainly the development of FFAG's).
\section{The target issue: the HARP experiment at CERN PS.}
\begin{figure}[htb]
\begin{center}
\vskip -1cm
\includegraphics[width=0.55\linewidth]{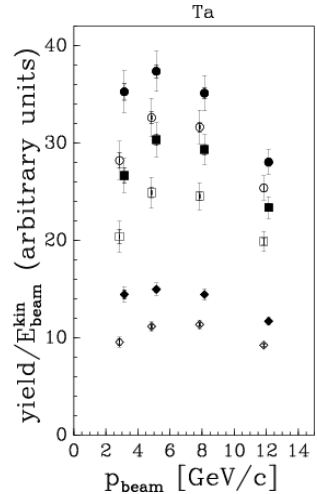}
\end{center}
\vskip -1cm
\caption{$\pi^+$ (closed symbols) and $\pi^-$
(open symbols) yields 
for different design of the NF focussing stage. The circles indicate the
integral over the full HARP acceptance, the squares are integrated
over 0.35 rad $\leq \theta \leq \ $ 0.95 rad, while the diamonds require in addition
the momentum cut 250 MeV/c $\leq p \leq $ 500 MeV/c.}
\label{fig:NF}
\end{figure}

The baseline option for a $\nu$F target is a Hg jet target with impinging
particles at energies $10 \pm 5 $ GeV. Available data are very scarce
and for the  tuning of the MC simulations of the $\nu$F beamline 
the HARP data on heavy targets, such as Ta or Pb,
are of utmost importance.
In the kinematics range of interest for a $\nu$F,
the pion yield increases linearly with momentum and has
an optimum between 5 GeV/c and 8 GeV/c, as shown  in
figure \ref{fig:NF}.
\begin{figure}[htb]
\begin{center}
\vskip -.5cm
\includegraphics[width=0.85\linewidth]{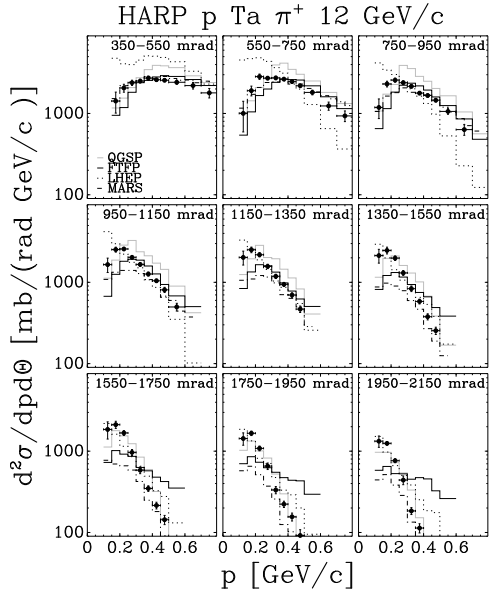}
\end{center}

\vskip -1.2cm
\caption{Experimental results from HARP at 12 GeV/c for p-Ta cross sections
for $\pi^{+}$ production, as compared to MC models, from ref. \cite{harp:LA}}
\label{fig:Ta}
\end{figure}
Final results for pion production on heavy targets
have been published in reference~\cite{harp:LA} and an example
of a comparison with available MC simulations is outlined in figure \ref{fig:Ta}.
None of the considered models describe fully HARP data. However,
$\pi^+$ production is described better than $\pi^-$ production.

In a $\nu$F,
the produced pions are then collected through a magnetic horn or focussed 
through a superconducting solenoid (baseline design).
 The MERIT (MERcury Intense Target) experiment at 
CERN~\cite{merit} has studied the feasibility of a mercury-jet target 
  for a 4 MW proton beam  with solenoidal pion capture, obtaining
positive results.

\section{The cooling issue: the MICE experiment at RAL.}
 The cooling of muons 
(accounting for $ \sim 20 \%$ of the final costs)
increases the performances of a $\nu$F up to a factor 10. Due to their short 
lifetime ($2.2 \mu$s),
 novel methods such as the
ionization cooling \cite{skrinsky} must be used. 
The cooling of the transverse phase-space coordinates of a muon
beam can be accomplished by passing it through an energy-absorbing material
and an accelerating structure, both embedded within a focusing magnetic 
lattice. Both longitudinal and transverse momentum are lost in the absorber
while the RF-cavities restore only the longitudinal component. 

\begin{figure}
\vskip -3cm
\begin{center}
\includegraphics[width=.90\linewidth]{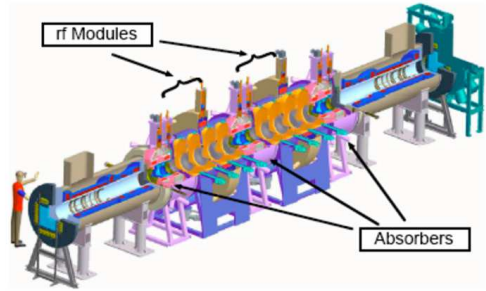}
\end{center}
\vskip -3.5cm
\caption{Layout of the MICE experiment at RAL. 
The secondary $\mu$ beam from ISIS enters 
from the lower left. The cooling section
is put between two magnetic spectrometers and two TOF stations
to measure particle parameters.}
\label{fig:mice}
\end{figure}

The MICE experiment \cite{mice} at RAL aims at a
systematic study of one cell of the US Feasibility Study 2 cooling channel (
see figure \ref{fig:mice} for its layout).
A secondary muon beam from ISIS (140-240 Mev/C central momentum) enters 
the cooling channel after a diffuser. Pions from a movable Ti target
grazing the primary ISIS beam, during its flat top, are captured by a
quadrupole triplet and then momentum selected. 
Muons from the following pion decays
inside a 5m long, 5 T decay solenoid are  momentum selected and
directed towards the MICE apparatus. 
The 5.5 m long cooling section cell consists of three low-Z absorbers 
and eight 201 MHz RF
cavities encircled by SC lattice solenoids, providing strong focussing.
While hydrogen absorbers are the best, a more practical absorption 
media may be He, solid LiH or Be. 
Particles are measured before and after the cooling section
by two magnetic spectrometers complemented by TOF detectors. 
For each particle the trackers determine x,y, 
x'=$p_x/p_z$,y'=$p_y/p_z$ and  t'=$E/p_z$ coordinates,
while the TOF stations measure the time coordinate t. 
For an ensemble of N particles, the
input and output emittances are thus measured  with high precision ($0.1 \%$),
at a level   
 not within reach of
conventional multiparticle methods.

The driving design criteria of the MICE beam instrumentation are robustness, 
in particular of the tracking detectors, to sustain the severe 
background conditions nearby the RFs and redundancy in PID 
in order to keep contaminations ($e, \pi$) well below $1 \%$.
Each magnetic spectrometer consists of a superconducting 4 T solenoid of 
40 cm bore, containing 5 planes of scintillating fiber detectors.
Each station 
is composed of three doublet layers in stereo view arrangement. 

Particle identification is obtained upstream the first solenoid by two 
TOF stations (TOF0/TOF1) and two aerogel Cherenkov counters (CKVa/CKVb).
$\pi/\mu$ separation is obtained via the Cherenkov counters for momenta
bigger than 210 MeV/c; below only the tof measurement is available.
Downstream the PID is obtained via a further TOF station (TOF2)
and an electromagnetic calorimeter (EMCAL). All downstream
detectors and the TOF1 station must be shielded against stray magnetic
fields from solenoids (up to 1000-1500 G with components along the PMT axis up to
400 G). While TOF1 will be shielded by a double-sided shielding cage
that fully contains the detector, TOF2 and EMCAL PMT's will be shielded 
locally by individual soft iron massive boxes.
The TOF stations share a common design based on fast scintillator counters
along X/Y directions (to increase measurement redundancy) read at both
edges by fast R4998 Hamamatsu photomultipliers ($\sim 160$ ps TTS, 0.7 ns 
risetime).   
A coincidence with TOF2 will select particles traversing the entire cooling 
channel. In addition the use of 
an electromagnetic calorimeter (EMCAL) will help to distinguish 
the genuine variation of emittance due to cooling from the one due to 
losses and $\mu \mapsto e$ decays.

The EMCAL is a Pb-scintillating fiber calorimeter (KL), of the KLOE type
\cite{kloe}, with 1-mm diameter blue scintillating fibers glued between
0.3 mm thick grooved lead plates to be followed by a muon ranger (SW), made
of a $1 \ m^3$ fully sensitive segmented scintillator block. 
This ``spaghetti'' design for KL offers the possibility of a fine sampling and 
optimal lateral uniformity. 
Both TOF (INFN MIB) and KL (INFN RM3) prototypes have been 
tested in the Frascati
BTF testbeam with satisfactory results. 
As an example, the TOF counters intrinsic resolution was around 50 ps.
Up to now, only the upstream 
PID detectors and KL are installed at RAL. 

MICE will be accomplished in steps during two phases:
first to characterize the incoming beam and demonstrating the capability
to do a high precision measure of emittance (PHASE I) and then 
to measure the transverse cooling 
for a variety of experimental situations (PHASE II). 
\section{Conclusions}
\noindent

Experimental R\&D results may soon strengthen the physics case for
a $\nu$F. Establishing the key techniques by the end of this decade,
can pave the way to build a facility in the next one.


\end{document}